# Shell Structure of Exotic Nuclei

J. Dobaczewski,[1] N. Michel,[2,3] W. Nazarewicz,[1–3] M. Płoszajczak,[4] J. Rotureau[2,3]

[1]Institute of Theoretical Physics, Warsaw University
ul. Hoża 69, 00-681 Warsaw, Poland
[2]Department of Physics & Astronomy, University of Tennessee
Knoxville, Tennessee 37996, USA
[3]Physics Division, Oak Ridge National Laboratory
P.O. Box 2008, Oak Ridge, Tennessee 37831, USA
[4]Grand Accélérateur National d'Ions Lourds (GANIL)
CEA/DSM - CNRS/IN2P3, BP 55027, F-14076 Caen Cedex, France

October 31, 2018

## Abstract

Theoretical predictions and experimental discoveries for neutron-rich, short-lived nuclei far from stability indicate that the familiar concept of nucleonic shell structure should be considered as less robust than previously thought. The notion of single-particle motion in exotic nuclei is reviewed with a particular focus on three aspects: (i) variations of nuclear mean field with neutron excess due to tensor interactions; (ii) importance of many-body correlations; and (iii) influence of open channels on properties of weakly bound and unbound nuclear states.

## 1 Introduction

Shell structure is a fundamental property of finite Fermi systems and atomic nuclei are not exceptions. In fact, the very concept of single-particle motion is a cornerstone of nuclear structure [1]. The stability of nuclei is rooted in non-uniformities of the single-particle level distribution and presence of magic gaps, and those can be traced back to classical periodic orbits of a nucleon moving in a one-body potential [2]. The shell effects are, therefore, intimately related to the mean-field approximation, to which the very notion of individual particle orbits is inherent.

The average potential of the nuclear Shell Model (SM), characterized by a flat bottom, a relatively narrow surface region, and a strong spin-orbit term, was introduced in 1949 by Maria Göppert-Mayer [3], Otto Haxel, Hans Jensen, and Hans Suess [4]. This SM concept, illustrated in the left-hand-side diagram of Fig. 1, explains the nuclear behavior in terms of single nucleons moving in individual orbits. But how robust is this picture? In nuclei close to the beta stability line, the modern nuclear SM, in which the dominant one-body behavior is augmented by a two-body residual interaction, is a very powerful tool [7]. However, a significant new theme concerns shell structure near the particle drip lines and in the superheavy nuclei. Theoretical predictions and experimental discoveries in the last decade indicate that nucleonic shell structure is being recognized now as a more local concept [8, 9, 10]. The experimental data indicate that the magic numbers in neutron-rich nuclei are not the immutable benchmarks they were once thought to be [11]: the magic numbers at $N$=20 and 28 fade away with neutron number and the new magic numbers at $N$=14, 16, and 32 seem to appear.

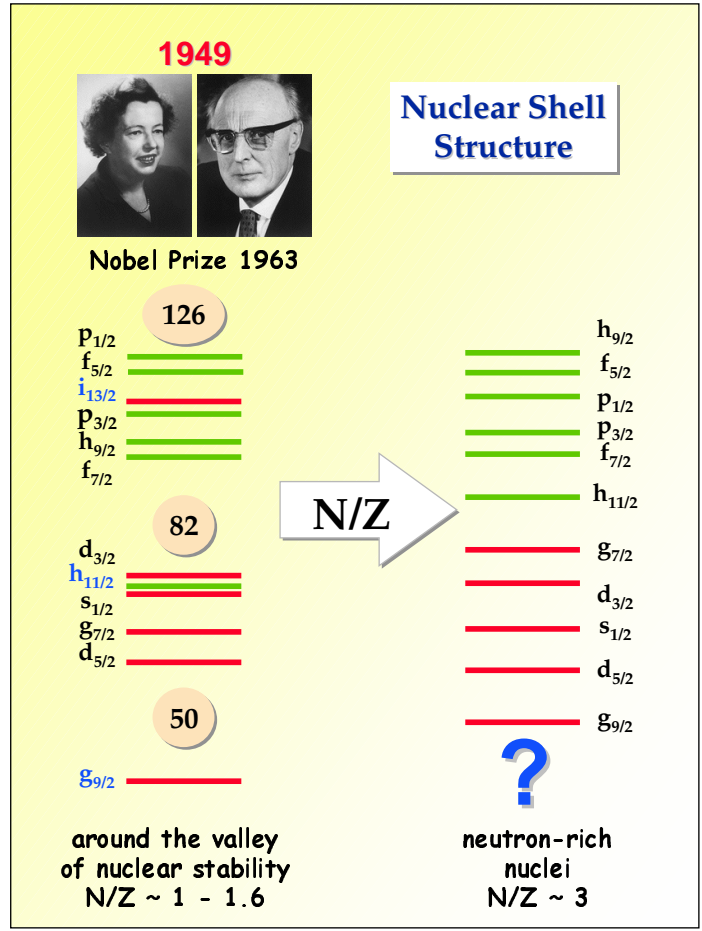

Figure 1: The cornerstone of nuclear structure for over half a century has been the shell model of Göppert-Mayer and Jensen, in which each nucleon is assumed to move in average potential. The left-hand-side diagram shows the shell structure characteristic of nuclei close to the valley of stability. The combined effects of flat bottom and small diffuseness of the mean-field potential, together with a strong spin-orbit term, result in a sizeable angular-momentum splitting of single-particle levels that yields the experimentally observed shell and subshell closures. The right-hand-side diagram shows schematically the shell structure that may exist in neutron-rich nuclei, which corresponds to a more uniform distribution of energy levels and the quenching of known magic gaps. (Based on Refs. [5, 6].)

Why is shell structure changing in the neutron-rich environment? There are several good reasons for this. *First*, as discussed in Sec. 2, the nuclear mean field is expected to strongly depend on the orbits being filled due to the tensor components of effective interaction. *Second*, many-body correlations, such as pairing, involving weakly bound and unbound nucleons become crucial when the neutron separation energy $S_n$ is small. Indeed, as seen from the approximate relation between the neutron Fermi level, $\lambda_n$, neutron pairing gap, $\Delta_n$, and $S_n$ [12, 6]:

$$S_n \approx -\lambda_n - \Delta_n, \qquad (1)$$

in the limit of weak binding, the single-particle field characterized by $\lambda_n$ and many-body correlations represented by the pairing field $\Delta_n$ become equally important. In other words, contrary to the situation encountered close to the line of beta stability, configuration-mixing effects in loosely bound nuclei near the drip lines can no longer be treated as a small perturbation atop the dominant mean field; they are of primary importance for the very existence of these systems. This means that in the limit of small $S_n$ the notion of a single-particle motion in a mean field, the basic idea behind the nuclear SM, is no longer a viable ansatz. *Third*, the nucleus is an open quantum system (OQS). The presence of states that are unbound to particle-emission may have significant impact on spectroscopic properties of nuclei, especially those close to the particle drip lines. In its standard realization [7, 13], the nuclear shell model assumes that the many-nucleon system is perfectly isolated from an external environment of scattering states and decay channels. The validity of such a closed quantum system (CQS) framework is sometimes justified by relatively high one-particle (neutron or proton) separation energies in nuclei close to the valley of beta stability. However, weakly bound or unbound nuclear states cannot be treated in a CQS formalism. As discussed in Sec. 3, a consistent description of the interplay between scattering states, resonances, and bound states in the many-body wave function requires an OQS formulation. Figure 2 schematically illustrates different physics components that are important for the structure of neutron-rich nuclei.

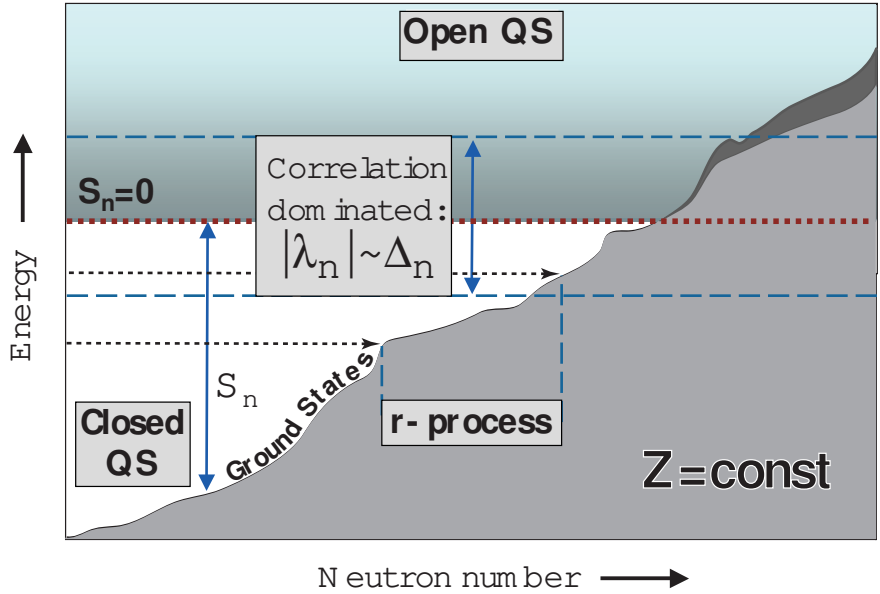

Figure 2: Schematic diagram illustrating various aspects of physics important in neutron-rich nuclei. One-neutron separation energies $S_n$, relative to the one-neutron drip-line limit ($S_n$=0), are shown for some isotopic chain as a function of $N$. The unbound nuclei beyond the one-neutron drip line are resonances. Their widths (represented by a dark area) vary depending on excitation energy and angular momentum. At low excitation energies, well-bound nuclei can be considered as closed quantum systems (QS). Weakly bound and unbound nuclei are open quantum systems that are strongly coupled to the environment of scattering and decay channels. The region of particularly strong many-body correlations, $S_n$<2 MeV or $|\lambda_n| \approx \Delta_n$ (1), is indicated. The astrophysical r-process is expected to proceed in the region of low separation energies, around 2-4 MeV [14]. In this region, effective interactions are strongly affected by isospin, and the many-body correlations and continuum effects are essential.

There are many examples of the impact of many-body correlations and continuum coupling on structural properties of neutron-rich nuclei. Halos with their low-energy decay thresholds and cluster structures are obvious cases [15]. The islands of inversions around magic numbers $N$=20 and 28 [16, 17, 18] offer another example. In those nuclei, dramatic asymmetry between proton and neutron Fermi surfaces gives rise to new couplings and the explanation involves coexistence effects due to intruder states and modifications of the effective interaction.

In the following sections, various aspects of physics responsible for structural (and spectroscopic) changes in exotic nuclei are briefly reviewed. Two detailed examples are given. Section 2 is devoted to effective interactions and shell structure in neutron-rich nuclei. The impact of the continuum coupling on unbound states in neutron-rich systems is discussed in Sec. 3.

## 2  Effective Interactions in Exotic Nuclei

The new experimental advances along the isospin axis and towards the territory of super-heavy elements at the limit of mass and charge require safe and reliable theoretical predictions of nuclear properties throughout the whole nuclear chart. The tool of choice is the nuclear density functional theory (DFT) based on the self-consistent Hartree-Fock-Bogoliubov (HFB) method. The key component is the universal energy density functional (EDF), which will be able to describe properties of finite nuclei as well as extended asymmetric nucleonic matter. The development of such a functional, including dynamical effects and symmetry restoration, is one of the main goals of the field [19, 20].

Developing a nuclear EDF requires a better understanding of the density and gradient dependence, spin effects, and pairing, as well as an improved treatment of symmetry-breaking effects and many-body correlations. Since the nuclear DFT deals with two kinds of nucleons, the isospin degree of freedom has to be introduced, and the isoscalar and isovector densities have to be considered [21]. A topic, relevant to the spin-isospin channel, that has recently become the focus of considerable attention is the influence of the tensor interaction on single-particle levels. It is discussed below.

### 2.1  Spin-Orbit Splitting and Tensor Interaction

Positions of single-particle levels change with varying neutron or proton numbers. These changes may lead to dramatic variations in properties of nuclei. For example, a shift in position of the neutron $1f_{5/2}$ level has been recently proposed to explain the appearance of a new $N=32$ magic gap in neutron-rich isotopes above $Z=20$. This is only one of several such examples recently identified in light nuclei and interpreted within the shell model in terms of a tensor interaction [22, 23]. However, the shell model is not an ideal tool to study single-particle energies, because it does not provide any description of the mean field of the nucleus.

Indeed, the single-particle states are theoretical constructs pertaining to the assumption that all nucleons move in a common, mean-field potential. Although such a mean-field picture of a nucleus gives its salient structural features, the actual nuclear state is certainly not a pure mean-field state. This can be understood in the spirit of the Kohn-Sham approach [24], in which mean-field orbitals represent correlated many-fermion states. The question of how much and what kind of correlations can be incorporated in a phenomenological EDF is still under debate. Certainly, the final comparison with experimental data should not be done by looking at the single-particle energies themselves. It is rather unclear how to reliably extract these theoretical objects from measured properties, especially in open-shell systems. Rather one should attempt determining in theory these same observables that are measured in experiment, whereupon single-particle energies keep the meaning of auxiliary quantities that facilitate analyzing final calculated results. In the present study, we follow this strategy by comparing measured and calculated masses of odd-$A$ nuclei.

The main focus of our analysis is on the spin-orbit (SO) properties of nuclei. The origin of the large nuclear one-body SO term, introduced in 1949 [3, 4], is still a matter of debate. On a microscopic level, a significant part of the SO splitting comes from the two-body SO and tensor forces [25, 26, 27, 28], and there is also a significant contribution from three-body forces [29, 30].

Suggestions to study tensor interactions within the self-consistent mean-field approach were made a long time ago [31, 32] but the scarce experimental data available (i.e., fairly short isotopic/isotonic chains) did not provide sufficient sensitivity to adjust the related coupling constants. On a one-body level, variation of the strength of the phenomenological SO potential due to the tensor force (sensitive to the effect of spin saturation) was studied in Refs. [33, 34]. Stimulated by the new data on neutron-rich nuclei, it is only very recently that the self-consistent treatment of the effective tensor interactions has been reintroduced [35, 36, 37].

Momentum-dependent zero-range tensor [31, 32] and spin-orbit (SO) [38] two-body interactions have the form

$$\hat{V}_{Te} = \tfrac{1}{2}t_e\left[\hat{\boldsymbol{k}}' \cdot \hat{\mathsf{S}} \cdot \hat{\boldsymbol{k}}' + \hat{\boldsymbol{k}} \cdot \hat{\mathsf{S}} \cdot \hat{\boldsymbol{k}}\right], \qquad (2)$$

$$\hat{V}_{To} = t_o \hat{\boldsymbol{k}}' \cdot \hat{\mathsf{S}} \cdot \hat{\boldsymbol{k}}, \qquad (3)$$

$$\hat{V}_{SO} = iW_0 \hat{\boldsymbol{S}} \cdot \left[\hat{\boldsymbol{k}}' \times \hat{\boldsymbol{k}}\right], \qquad (4)$$

where the vector and tensor spin operators read

$$\hat{\boldsymbol{S}} = \boldsymbol{\sigma}_1 + \boldsymbol{\sigma}_2, \qquad (5)$$

$$\hat{\mathsf{S}}^{ij} = \tfrac{3}{2}\left[\sigma_1^i \sigma_2^j + \sigma_1^j \sigma_2^i\right] - \delta^{ij} \boldsymbol{\sigma}_1 \cdot \boldsymbol{\sigma}_2. \qquad (6)$$

When averaged with one-body density matrices, these interactions contribute to the following terms in the energy-density functional (EDF) (see Refs. [39, 21] for derivations),

$$\mathcal{H}_T = \tfrac{5}{8}\left[t_e \boldsymbol{J}_n \cdot \boldsymbol{J}_p + t_o(\boldsymbol{J}_0^2 - \boldsymbol{J}_n \cdot \boldsymbol{J}_p)\right], \qquad (7)$$

$$\mathcal{H}_{SO} = \tfrac{1}{4}\left[3W_0 \boldsymbol{J}_0 \cdot \boldsymbol{\nabla}\rho_0 + W_1 \boldsymbol{J}_1 \cdot \boldsymbol{\nabla}\rho_1\right], \qquad (8)$$

where $W_1 = W_0$ and the conservation of time-reversal and spherical symmetries was assumed. Here, $\rho_t$ and $\boldsymbol{J}_t$ are the neutron, proton, isoscalar, and isovector particle and vector SO densities [38, 39, 21] for $t=n$, $p$, 0, and 1, respectively. Within the EDF formalism, one extends the SO energy density (8) to the case of $W_1 \neq W_0$ [40].

Apart from the contribution of the SO energy density to the central potential, variation of the SO and tensor terms with respect to the densities yields the following form factors of the one-body SO potentials for neutrons and protons,

$$\boldsymbol{W}_n^{SO} = \tfrac{5t_e+5t_o}{8}\boldsymbol{J}_p + \tfrac{5t_o}{4}\boldsymbol{J}_n + \tfrac{3W_0-W_1}{4}\boldsymbol{\nabla}\rho_p + \tfrac{3W_0+W_1}{4}\boldsymbol{\nabla}\rho_n, \qquad (9)$$

$$\boldsymbol{W}_p^{SO} = \tfrac{5t_e+5t_o}{8}\boldsymbol{J}_n + \tfrac{5t_o}{4}\boldsymbol{J}_p + \tfrac{3W_0-W_1}{4}\boldsymbol{\nabla}\rho_n + \tfrac{3W_0+W_1}{4}\boldsymbol{\nabla}\rho_p. \qquad (10)$$

Hence, it is clear that the only spectroscopic effect of tensor interaction is a modification of the SO splitting of the single-particle levels. From the point of view of one-body properties, tensor interactions act similarly to two-body SO forces. However, the latter ones induce the SO splitting that is weakly dependent on shell filling. This is so because the corresponding form factors in Eqs. (9) and (10) are given by the radial derivatives of densities, $\boldsymbol{\nabla}\rho = \tfrac{\boldsymbol{r}}{r}\tfrac{d\rho}{dr}$, which are weakly dependent on shell effects. On the other hand, the SO splitting induced by the tensor forces depends strongly on the shell filling, because the corresponding form factors are given by the SO densities, $\boldsymbol{J} = \tfrac{\boldsymbol{r}}{r}J(r)$. Indeed, when only one of the SO partners is occupied (spin-unsaturated system), the SO density $J(r)$ is large, and when both partners are occupied (spin-saturated system), the SO density is small (see Ref. [35] for numerical examples).

Within the SM, these simple SO effects are called "level attraction" or "level repulsion" [22, 23], but in reality they result from characteristic changes of the SO mean fields. Indeed, although the tensor SO mean fields are dominated by contributions from high-$j$ orbitals, those of lower-$j$ orbitals are also important. Instead of interpreting the effects of tensor interaction in terms of the "level-level" interactions, a more physical explanation can be given in terms of a two-step argument. First, the SO densities depend on occupations of the SO partners, and, second, the SO partners are split according to the one-body SO mean fields given by form factors (9,10).

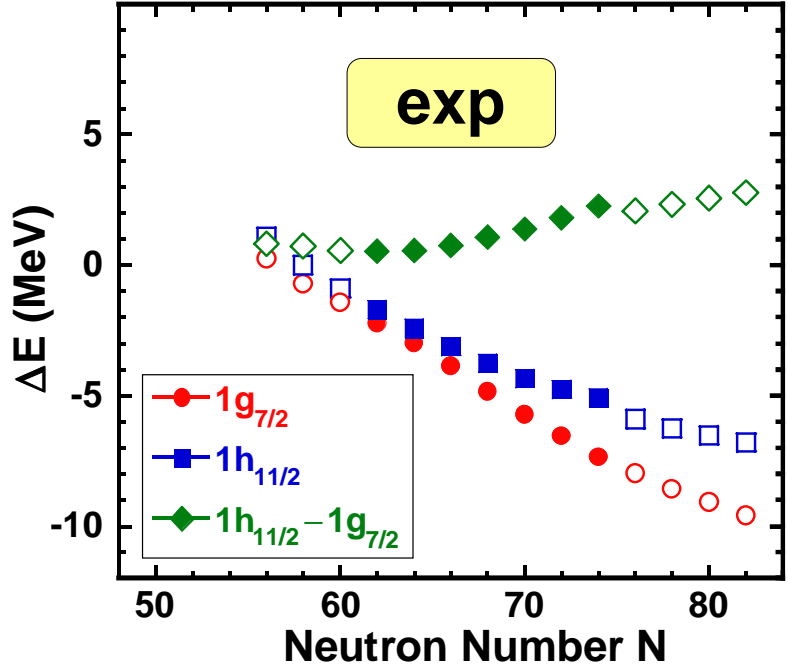

Figure 3: Experimental differences of 11/2− and 7/2+ energies in the $Z=51$ isotopes $E[1h_{11/2}]-E[1g_{7/2}]$ (diamonds) compared to differences $E[1h_{11/2}]-E[0^+]$ (squares) and $E[1g_{7/2}]-E[0^+]$ (circles) relative to the ground states of the $Z=50$ isotones. Open symbols show data for the $Z=51$ levels where there is no information from transfer reactions about the single-particle character of the states. (From Ref. [41].)

In the present study, we apply tensor terms in the EDF to describe relative changes of positions of the proton $1h_{11/2}$ and $1g_{7/2}$ levels. Figure 3 shows experimental differences $E[1h_{11/2}] - E[1g_{7/2}]$ in the $Z=51$ isotopes between $N=56$ and 82 [41]. The figure also shows differences $E[1h_{11/2}] - E[0^+]$ and $E[1g_{7/2}] - E[0^+]$ (circles) relative to the ground states of the $Z=50$ isotones. In each case, the symbol $E[\text{state}]$ denotes the *total* mass of the nucleus in the corresponding state.

Figure 4 shows analogous results obtained within the spherical HFB calculations performed for the Skyrme interactions SLy4 [42] and SkO [16]. The HFB solutions in the $Z=51$ isotopes were obtained by self-consistently blocking one proton in the $1h_{11/2}$ and $1g_{7/2}$ orbits, while the standard unblocked calculations were performed in the $Z=50$ isotones. Consequently, energy differences in Fig. 4 correspond to differences of the *total* calculated masses; thus they include the odd-particle polarization effects.

Without tensor terms (top panels), $E[1h_{11/2}] - E[1g_{7/2}]$ values are almost constant with $N$. This illustrates the particle-number dependence of the standard SO splitting, which is dictated by gradients of densities. This result does not depend on the type of coupling which is different for the two Skyrme interactions used. Indeed, for SLy4, one has $W_1 = W_0$, and hence the proton SO splitting couples twice as strong to the gradient of the proton density than to the gradient of the neutron density (cf. Eq. (10)). On the other hand, for SkO, one has $W_1 \simeq -W_0$ and the type of coupling is opposite – mostly to the gradient of the neutron density. Nevertheless, for both interactions, the standard SO splitting is almost constant.

In the present study, we show results for tensor interaction $t_o = 0$ and $t_e = 200\,\text{MeV}\,\text{fm}^5$, proposed in Ref. [35]. When such tensor interaction is taken into account, the results are qualitatively different. Namely, differences $E[1h_{11/2}] - E[1g_{7/2}]$ are smallest near $N=70$ where the neutron SO densities are smallest. (The neutron $1g_{7/2}$ shell is almost filled while the neutron $1h_{11/2}$ shell is almost empty – i.e., the system is close to being spin-saturated.) On both sides of $N=70$, differences $E[1h_{11/2}] - E[1g_{7/2}]$ increase; for $N < 70$ ($N < 70$) this is due to decreasing (increasing) occupations of the $1g_{7/2}$ ($1h_{11/2}$) shells. Therefore, on both sides, the neutron SO densities increase; hence, the $1h$ and $1g$ SO splittings decrease, resulting in the observed pattern of differences $E[1h_{11/2}] - E[1g_{7/2}]$.

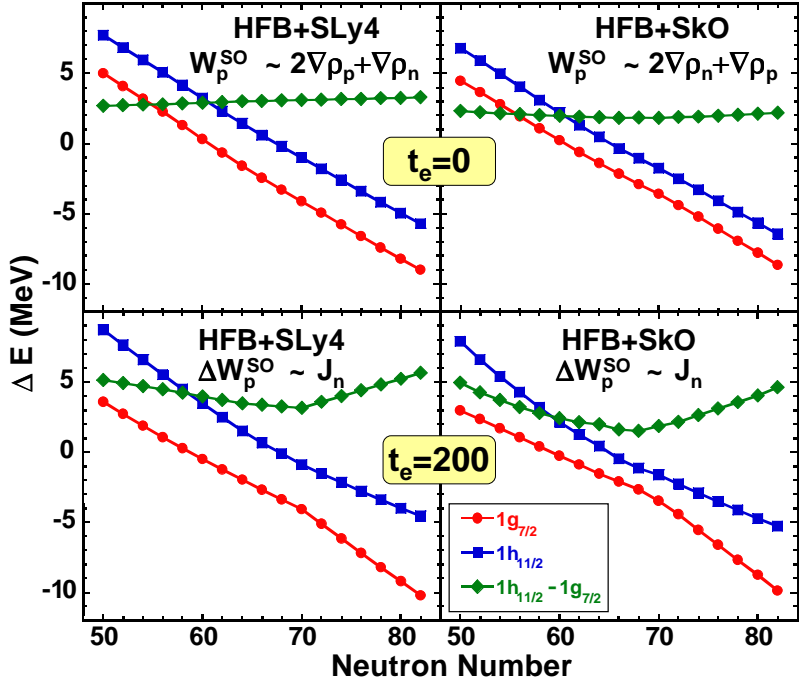

Figure 4: Differences of total calculated energies in the $Z=51$ isotopes, analogous to those shown in Fig. 3. Top panels show results obtained without tensor terms, $t_o = t_e = 0$, while bottom panels correspond to tensor-even terms of $t_e = 200\,\text{MeV}\,\text{fm}^5$. In each case, we indicate the type of SO coupling within the proton standard SO ($W_p^{SO}$) and tensor SO ($\Delta W_p^{SO}$) part of the form factor (10). Left and right panels show results obtained with the SLy4 [42] and SkO [16] interactions, respectively.

The calculated pattern of the $1h_{11/2}$–$1g_{7/2}$ splitting reflects the experimental trends; however, in experiment the minimum splitting is close to $N=64$, and the increase of splitting at $N<64$ is quite weak. In order to confirm the observed pattern, it would be crucial to identify the proton $1h_{11/2}$ and $1g_{7/2}$ levels in the $^{101-105}$Sb isotopes. Moreover, better Skyrme parametrizations are needed to shift the minimum from $N=70$ towards $N=64$ by shifting the neutron $3s_{1/2}$ and $2d_{3/2}$ levels up, closer to the $1h_{11/2}$ level. We note here in passing that in Ref. [36] such a shift was done artificially, without using the real self-consistent energies of the Gogny interaction.

## 3 Particle Continuum and Shell Structure

The many-body nuclear Hamiltonian does not describe just one nucleus $(N, Z)$, but all nuclei that can exist. In this sense, a nucleus is never isolated (closed) but communicates with other nuclei through decays and captures. If the continuum space is not considered, this communication is not allowed:

the system is closed. A consistent description of the interplay between scattering states, resonances, and bound states in the many-body wave function requires an OQS formulation (see Refs. [43] and references quoted therein). Properties of unbound states lying above the particle (or cluster) threshold directly impact the continuum structure. Coupling to the particle continuum is also important for weakly bound states, such as halos. A classic example of a threshold effect is the Thomas-Ehrman shift [44, 45] shown in Fig. 5 (left) which manifests itself in the striking asymmetry in the energy spectra between mirror nuclei having different particle emission thresholds. As argued by Ikeda *et al.* [46], clustering in nuclei becomes relevant for states in nuclei close to their cluster decay thresholds. This is a particular aspect of a general phenomenon of the alignment of near-threshold states with the decay channel which finds its explanation in generic features of the continuum coupling close to the threshold [43, 47, 48, 49]. The right-hand-side of Fig. 5 shows the example of the phenomenon of the alignment of the weakly bound $0_2^+$ state in $^{16}$O at 6049 keV with the $^{12}$C+$\alpha$ decay channel (threshold at 7162 keV). Another splendid example is the $0_2^+$ resonance in $^{12}$C at 7.65 MeV that lies slightly above the triple-alpha threshold at 7.275 MeV. Both states, crucial for our understanding of stellar nucleosynthesis, have an extremely complex character in the language of standard SM. Their description requires proper treatment of many-body correlations and continuum.

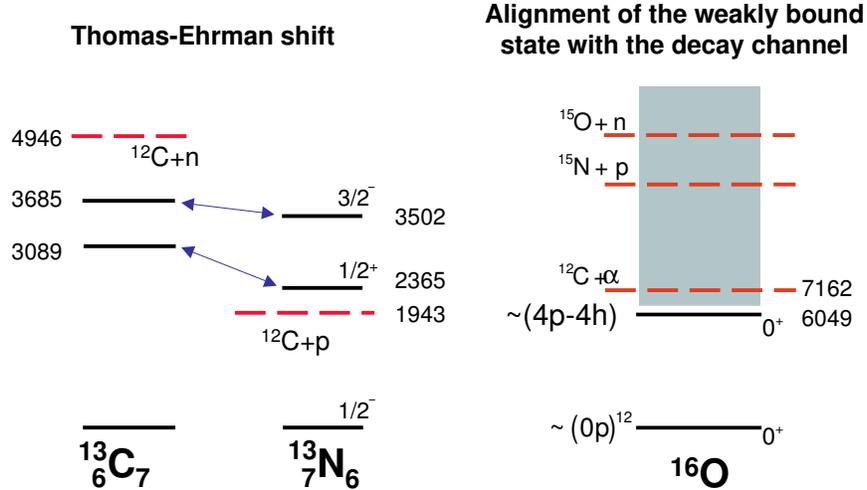

Figure 5: Examples of the continuum coupling on spectroscopic properties of nuclei. Left: the Thomas-Ehrman shift between $1/2^+$ and $3/2^-$ states in mirror nuclei $^{13}$N (one-proton threshold at 1943 keV) and $^{13}$C (one-neutron threshold at 4946 keV). Right: the phenomenon of the alignment of the weakly bound state with the cluster decay channel; the appearance of the $0_2^+$ state in $^{16}$O slightly below the $^{12}$C+$\alpha$ threshold.

The mechanism of alignment of bound and unbound near-threshold states with the decay channel has its microscopic origin in the continuum coupling which is anomalously strong in the neighborhood of the particle-emission threshold [43]. The continuum-coupling energy correction to the CQS eigenvalue exhibits a cusp behavior [43, 48, 49] near the threshold which reflects the change in the configuration mixing due to the *external coupling* to the decay channel(s). This external coupling is responsible for an additional configuration mixing (the so-called *external mixing*), which tends to increase a similarity with the neighboring decay channel for near-threshold states. The mechanism of *channel alignment* is generic in a sense that it does not depend both on single-particle energies/resonances (the single-particle potential) and on any detailed features of the two-body interaction. Therefore the appearance of cluster states near cluster-emission threshold, like the appearance of halo-states near the corresponding one- and two-neutron emission thresholds, is not an accidental phenomenon but a generic feature of any correlated OQS close to the particle/cluster-emission thresholds.

The effect of resonances and the non-resonant scattering states can be considered in the open quantum system (OQS) extension of the SM, the so-called continuum SM [43]. In this work, we apply the complex-energy implementation of the continuum SM, the so-called Gamow Shell Model (GSM) [50, 51] in the version of Refs. [52, 53]. GSM is a multi-configurational SM with a single-particle (s.p.) basis given by the Berggren ensemble [54]. The GSM can be viewed as a quasi-stationary many-body OQS formalism.

## 3.1 The Gamow Shell Model

In the roots of GSM lies the Berggren one-body completeness relation [54, 55, 56] that provides a mathematical foundation for unifying bound and unbound states. The Berggren ensemble consists of bound, resonant, and scattering s.p. wave functions generated by a finite-depth potential $V(r)$. The wave functions are regular solutions of the s.p. Schrödinger equation,

$$u''_{\mathcal{B}}(r) = \left[\frac{\ell(\ell+1)}{r^2} + \frac{2m}{\hbar^2}V(r) - k^2\right] u_{\mathcal{B}}(r), \quad (11)$$

that obey outgoing or scattering boundary conditions:

$$u_{\mathcal{B}}(r)_{r\to+\infty} \sim C_+ \, H_\ell^+(kr), \quad (12)$$

where $k$ and $\ell$ are the s.p. linear and angular momentum, respectively. For the resonant states, $C_-=0$ (outgoing boundary condition). In this paper, as we consider only valence neutrons, $H_\ell^\pm$ is a Hankel wave function. Normalization constants $C_+$ and $C_-$ are determined from the condition that the radial wave functions $u(r)$ are normalized to unity (for resonant states) or to the Dirac delta (for scattering states).

For a given partial wave $(\ell, j)$, the scattering states are distributed along the contour $L_+^{\ell_j}$ in the complex momentum plane. The set $|u_{\mathcal{B}}\rangle$ of all bound and resonant states enclosed between $L_+^{\ell_j}$ and the real $k$-axis, and scattering states is complete [54, 55, 56]:

$$\sum_{\mathcal{B}} \kern-1.2em\int \; |u_{\mathcal{B}}\rangle\langle\widetilde{u_{\mathcal{B}}}| = 1. \quad (13)$$

The many-body Berggren basis is that of Slater determinants built from s.p. states of Eqs. (12,13). In the Berggren representation, the SM Hamiltonian becomes a complex symmetric matrix. The fundamental difference between GSM and the real energy SM is that the many-body resonant states of the GSM are embedded in the background of scattering eigenstates, so that one needs a criterion to isolate them. The overlap method introduced in Ref. [52] has proven to be very efficient in this respect.

## 3.2 Neutron-rich He Isotopes

Coupling to the non-resonant scattering continuum may change significantly energies of the many-body states close to the particle-emission threshold [43]. In this section, we shall examine the interplay between the continuum coupling and SO interaction in $^5$He and $^7$He. The GSM Hamiltonian consists of a one-body Woods-Saxon (WS) potential and a two-body residual interaction:

$$H = \sum_i \left[\frac{p_i^2}{2m} + V_{WS}(r_i)\right] + V_{res}^{(2)}. \qquad (14)$$

The WS potential contains the central term and the SO term:

$$V(r) = -V_{WS} \cdot f(r) - 4\, V_{so}\,(\boldsymbol{l}\cdot\boldsymbol{s})\,\frac{1}{r}\left|\frac{df(r)}{dr}\right|, \qquad (15)$$

where

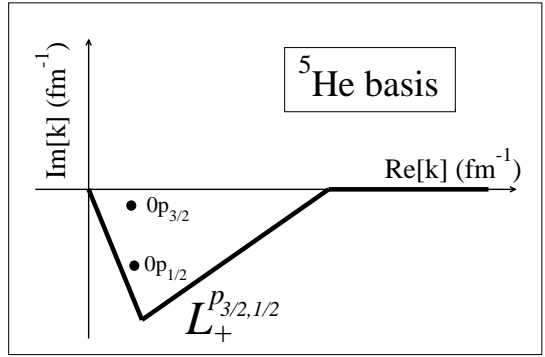

Figure 6: The Berggren ensemble in the complex $k$-plane employed in this work for the helium isotopes. The S-matrix poles ($0p_{3/2}$ and $0p_{1/2}$ and Gamow states) are denoted by dots. Scattering ($p_{3/2}$ and $p_{1/2}$) contours are shown by a thick solid line.

$$f(r) = \left[1 + \exp\left(\frac{r - R_0}{d}\right)\right]^{-1} \qquad (16)$$

is the usual WS form factor characterized by radius $R_0$, diffuseness $d$, depth $V_{WS}$, and SO strength $V_{so}$. We employed the "$^5$He" WS parameter set [52] which reproduces the experimental energies and widths of known s.p. resonances $3/2_1^-$ and $1/2_1^-$ in $^5$He ($0p_{3/2}$ and $0p_{1/2}$ resonant states in our model). For a residual interaction, we took the finite-range surface Gaussian interaction (SGI) [53]

$$V_{res}^{(2)} = \sum_{i<j} V_0^{(J)} \cdot \exp\left(-\left[\frac{\boldsymbol{r}_i - \boldsymbol{r}_j}{\mu}\right]^2\right) \cdot \delta(|\boldsymbol{r}_i| + |\boldsymbol{r}_j| - 2R_0) \qquad (17)$$

with the range $\mu$=1 fm and the coupling constants depending on the total angular momentum $J$ of the neutron pair: $V_0^{(0)} = -403$ MeV fm$^3$, $V_0^{(2)} = -392$ MeV fm$^3$. These constants are fitted to reproduce the ground-state (g.s.) binding energies of $^6$He and $^7$He in GSM.

Figure 6 shows the Berggren ensemble employed in this work. The doubly magic nucleus $^4$He has been assumed as an inert core. The valence space for neutrons consists of the $0p_{3/2}$ and $0p_{1/2}$ resonant states, and complex-momentum $p_{3/2}$ and $p_{1/2}$ scattering continua. For both $L_+^{\ell j}$ contours of Fig. 6, the maximal value for $k$ is $k_{max}$ =3.27 fm$^{-1}$. The contours have been discretized with up to 27 points. The discretization points and normalizations of the corresponding discretized scattering states have been chosen according to the Gauss-Legendre integration formula The GSM eigenvalues for $^7$He have been found using the DMRG technique described in Ref. [57].

In order to illuminate certain aspects of GSM results (e.g., non-resonant continuum coupling or configuration mixing), we have introduced two simplified SM schemes: the GSM in the pole approximation (GSM-p) and the harmonic oscillator shell model (HO-SM). In GSM-p, the scattering components of the shell-model space are disregarded, and only the $0p_{3/2}$ and $0p_{1/2}$ resonant states are present in the basis. In this case, the one-body completeness relation (13) is obviously violated. Still, the comparison between full GSM results and GSM-p is instructive, as it illustrates the influence of the non-resonant continuum subspace on the GSM results. For the HO-SM calculation, the radial wave functions of the GSM-p basis are replaced by those of the spherical harmonic oscillator (HO) with the frequency $\hbar\omega = 41A^{-1/3}$ MeV. The real energies of the resonant states define the one-body part of the HO-SM Hamiltonian. The HO-SM scheme is supposed to illustrate the "standard" CQS SM calculations in which only bound valence shells are considered in the s.p. basis. The energy difference $\Delta E = E_{1/2_1^-} - E_{3/2_1^-}$ of the two lowest states in $^5$He and $^7$He, calculated as a function of the SO term in the one-body potential, is shown in Fig. (7). In $^5$He, these two states in our model space are the s.p. Gamow states $0p_{3/2}$ and $0p_{1/2}$, so that their energy difference is due to a one-body SO splitting. In the absence of the residual interaction, the same splitting is obviously obtained for $^7$He.

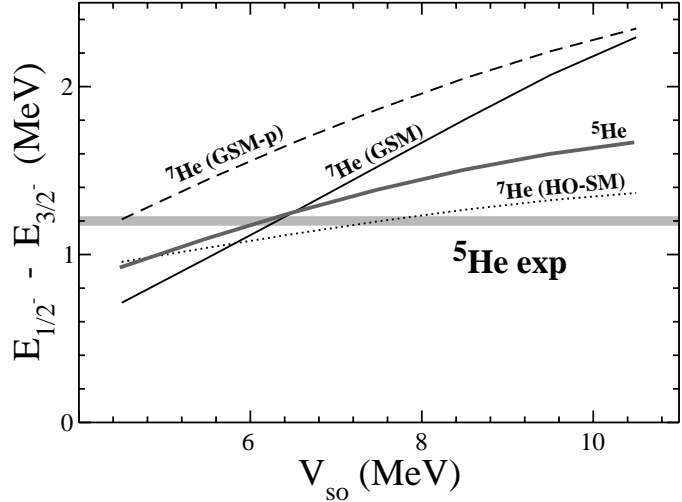

Figure 7: The energy difference, $\Delta E$, of $1/2_1^-$ and $3/2_1^-$ states in $^5$He and $^7$He as a function of the SO strength. The gray line shows $\Delta E$ in $^5$He. In this case $\Delta E$ is equivalent to a s.p. SO splitting. The full GSM result for $^7$He is given by the black solid line, and dotted and dashed lines mark HO-SM and GSM-p results, respectively. Experimental splitting in $^5$He is indicated.

To investigate the importance of the non-resonant continuum, we have performed the complete GSM calculation in the full model space, as well as the GSM-p and HO-SM calculations. In the pole approximation, $\Delta E$ follows closely the $^5$He result, albeit it is slightly larger. The calculated GSM-p states in $^7$He are almost the same as in the absence of the residual interaction, so the $^5$He and $^7$He curves differ only by a small shift due to the residual interaction. In the HO-SM variant, the residual interaction gives rise to a configuration mixing, and $\Delta E$ in $^7$He is slightly reduced as compared to the original SO splitting in $^5$He. In the full GSM calculations, the coupling to the non-resonant continuum comes to the fore. The energy difference in $^7$He varies steeply with $V_{\text{so}}$, contrary to other curves. The difference between GSM and GSM-p results reflects different correlation energies in the $1/2_1^-$ and $3/2_1^-$ states due to a coupling to the non-resonant continuum. This means that effects of the one-body SO interaction are effectively modified by the structure of loosely bound or unbound states, so that the same interaction can lead to a different value of $\Delta E$ in neighboring nuclei.

# 4 Conclusions

In this paper, we attempted to briefly address two questions lying at the heart of the radioactive nuclear beam research: (i) Why is the shell structure changing in exotic nuclei? (ii) Can one always define shell structure in the nuclear *terra incognita*?

The answers to these questions are not simple, as the underlying problem has many dimensions. Shell structure is a mean-field feature. If a mean field cannot be properly defined, the concept of an average potential is not useful. By now, we have learned that the structure of weakly bound, neutron-rich nuclei is strongly impacted by many-body correlations and coupling to open channels. In those exotic systems, such as halos and clusters, the picture of protons and neutrons moving independently in well-localized SM orbits is not appropriate. Another limit of the nuclear shell model is reached in the region of superheavy and hyperheavy nuclei [58]. Here, the Coulomb interaction cannot be treated as a small perturbation; it does influence significantly the nuclear mean field. In particular, shell closures are not expected to be robust in superheavy nuclei. Indeed, the theory predicts that beyond $Z=82$ and $N=126$ the familiar localization of shell effects at magic numbers is basically gone [59]. Moreover, due to the large *Coulomb frustration*, one expects the formation of voids in nuclei with very large atomic numbers. Preliminary calculations suggest that the mean fields in underlying exotic (bubble, toroidal, band-like) configurations are close in energy. What is the stability of these states? How strong is the coupling between the near-lying minima? A similar question has been asked in the context on diluted nuclear matter in the neutron star crust. Self-consistent calculations confirm the fact that the "pasta phase" might have a rather complex structure, various shapes can coexist, and the neutron star crust could be on the verge of a disordered phase [60, 61].

If the mean field (and shell structure) can be defined, changes of spectroscopic properties in exotic nuclei can be traced back to the three main factors: isospin dependence of effective interactions, symmetry-breaking effects and configuration mixing, and the continuum coupling. Usually, all these mechanisms are in play. This poses a serious challenge for nuclear structure theory and makes this area of research both demanding and fascinating.

## Acknowledgements

This work was supported in part by the U.S. Department of Energy under Contract Nos. DE-FG02-96ER40963 (University of Tennessee), DE-AC05-00OR22725 with UT-Battelle, LLC (Oak Ridge National Laboratory), DE-FG05-87ER40361 (Joint Institute for Heavy Ion Research), DE-FC02-07ER41457 (UNEDF, SciDAC-2); by the Polish Committee for Scientific Research (KBN) under Contract No. 1 P03B 059 27; and by the Foundation for Polish Science (FNP).